\documentstyle[12pt]{article}
\newcommand{\sv}[1]{{\mbox{\boldmath $#1$}}}
\newfont{\msbm}{msbm10}

\title{The dynamic effect of quantum decoherence.}

\author
{{L.V.Il'ichov}
\thanks{
e-mail: shalagin@iae.nsk.su;
fax: (383-2)333-863}  \\
\normalsize{\it Institute of Automation and Electrometry SB RAS,}\\
\normalsize{\it Novosibirsk State University,}\\
\normalsize{\it Novosibirsk 630090, Russian Federation}}

\date{}

\begin{document}

 %Inserted by TeXtelmExtel

\maketitle
%..........................................................

 %Inserted by TeXtelmExtel

 %Inserted by TeXtelmExtel

\begin{abstract}
        An effective force induced by spatially depending decoherence is predicted. The phenomenon is illustrated by the simple model of a 1/2-spin particle subjected to distributed unselective measurement of noncommuting spin components. 
\end{abstract}    

 %Inserted by TeXtelmExtel

%..........................................................

 %Inserted by TeXtelmExtel

 %Inserted by TeXtelmExtel

 %Inserted by TeXtelmExtel

 %Inserted by TeXtelmExtel

\newpage     
	The behavior of open quantum systems is extremely intriguing problem both from theoretical and experimental view points. There is a permanently extending class of phenomena where decoherent action of environment changes dramatically the evolution of quantum systems. These are the process of enantiomer conversion [1] and the conversion of nuclear spin isomers [2], decay of neutral kaons [3], neutrino oscillations [4], the so-called ``quantum Zeno effect'' [5] and its modification known as ``interaction-free measurement'' [6], etc. In these phenomena the influence of environment appears not as merely an annoying unavoidable factor, but as an indispensable ingredient.

 %Inserted by TeXtelmExtel

	The action of environment distinguishes a specific basis in the Hilbert space ${\cal H}$ of the considered quantum system -- the {\it pointer basis} (see, e.g. [7]). If the environment should appear to be the only factor of the system's evolution any coherence between different elements of the pointer basis would have disappeared. Thus all corresponding off-diagonal elements of the system's density matrix $\hat{\varrho}$ would have come to zero. Another particular basis in ${\cal H}$, which can be associated with the system, diagonalizes its Hamiltonian. The common key point of all the processes mentioned above is a discrepancy between these two basises. Manipulating the discrepancy, one can efficiently modify the evolution of the system. An example of such a manipulation is the so-called ``Sisiphus effect'' [8] -- a possible way of laser cooling of resonant atoms. In this case the pointer basis is the direct product of ``bare'' atom states and  photon-number states of quantized laser field. This basis does not coincides with the set of stationary states of ``dressed'' atom (the composit ``atom + laser field'' system). The second basis depends on the spatial configuration of the laser field in contrast to the pointer basis. This ${\sv r}$-dependence is of principal importance for the ``Sisiphus cooling''.

 %Inserted by TeXtelmExtel

	In the present work an opposite situation is considered -- the case of local $\sv r$-dependent pointer basis. I argue that under certain conditions this leads to effective forces acting on the system. To illustrate the phenomenon the simplest model is used -- a quantum particle with an internal 2D complex Hilbert space - the spin space - and three translational degrees of freedom. The notion of ``spin'' should be considered here in an extended sense - this may be the ordinary spin or equally the quantity associated with a dichotomous observable such as chirality. We assume that to every $\sv r$-point a local orthonormalized spin basis $|\psi_{+}(\sv r)\rangle$ and $|\psi_{-}(\sv r)\rangle$ is attached. Upon the fixation of a laboratory spin basis the states $|\psi_{\pm}(\sv r)\rangle$ can be identified with $\sv r$-dependent spinors -- 2-columns of complex numbers. The projectors onto these states can be expand in terms of three Pauli matrixes $\hat{\sigma}_{\alpha} (\alpha = 1, 2, 3)$:
\begin{equation}
{\hat P}_{\pm}(\sv r) \equiv |\psi_{\pm}(\sv r)\rangle\langle\psi_{\pm}(\sv r)| = \frac{1}{2}
\bigg(1 \pm {\vec n}(\sv r)\hat{\vec\sigma}\bigg), 
\label{2}
\end{equation}
where ${\vec n}(\sv r)$ is a unit vector in an effective 3D space associated with spin states.   
The Cartesian coordinate system of the 3D space is specified by the laboratory spin basis.
Any measurement of spin projection along ${\vec n}(\sv r)$ gives 1/2 for $|\psi_{+}(\sv r)\rangle$ and -1/2 for $|\psi_{-}(\sv r)\rangle$. Hence, one may use the following more convenient notations: 
\begin{equation}
|\psi_{+}(\sv r)\rangle \equiv |{\vec n}(\sv r)\rangle, \;\;\;\;
|\psi_{-}(\sv r)\rangle \equiv |-{\vec n}(\sv r)\rangle,
\label{1}
\end{equation}  
 
 %Inserted by TeXtelmExtel
 
	We advance now directly to the equation of the model:
\begin{equation}
\partial_{t}\hat{\varrho} + \frac{i}{2m}[{\hat{\sv p}}^2, \hat{\varrho}] =
\nu\bigg({\hat P}_{+}(\hat{\sv r}){\hat\varrho}{\hat P}_{+}(\hat{\sv r}) +
{\hat P}_{-}(\hat{\sv r}){\hat\varrho}{\hat P}_{-}(\hat{\sv r}) - {\hat\varrho}\bigg).
\label{3}
\end{equation}
The equation shows that the motion of the particle is accompanied by a decoherence process governed by the Lindblad form (the RHS of Eq.(\ref{3})). Locally this process looks like	
unselective spin measurements along ${\vec n}(\sv r)$. The measurements take place
throughout the space randomly in time with the average frequency $\nu$. It is important that in accordance
with the considered model the spin measurements are attended by no spatial localization. Hence this setting may provide the information on spin projection along $\vec n$ but not on the location of the measurement event. We prefer not to speak here on technical realizations of such a model and simply assume its possibility. One may alternatively assume the hypothetical natural origin of the RHS of Eq.(\ref{3}) in much the same way as Ghirardi, Rimini, and Weber do with their spontaneous localization model [9].  

 %Inserted by TeXtelmExtel

	Likewise Eq.(\ref{2}), one may write the density matrix in coordinate representation:
\begin{equation}
{\hat\varrho}(\sv r_1|\sv r_2) = \frac{1}{2}\bigg(\rho_0(\sv r_1|\sv r_2) + {\vec\rho}(\sv r_1|\sv r_2)\hat{\vec\sigma}\bigg). 
\label{4}
\end{equation}
The equations for the density $\rho_0(\sv r_1|\sv r_2)$ and for the orientation vector ${\vec\rho}(\sv r_1|\sv r_2)$ follow from Eq.(\ref{3}):
$$
\partial_t\rho_0(\sv r_1|\sv r_2) - \frac{i}{2m}(\Delta_{\sv r_1} - \Delta_{\sv r_2})\rho_0(\sv r_1|\sv r_2) 
$$
$$
= -\frac{\nu}{4}\bigg({\vec n}(\sv r_1) - {\vec n}(\sv r_2)\bigg)^2\rho_0(\sv r_1|\sv r_2) - 
\frac{i\nu}{2}\bigg({\vec n}(\sv r_1)\times{\vec n}(\sv r_2)\bigg)\cdot{\vec\rho}(\sv r_1|\sv r_2),
$$
\begin{equation}
\label{5}
\end{equation}
$$
\partial_t{\vec\rho}(\sv r_1|\sv r_2) - \frac{i}{2m}(\Delta_{\sv r_1} - 
\Delta_{\sv r_2}){\vec\rho}(\sv r_1|\sv r_2) 
$$
$$
= \frac{\nu}{2}{\vec n}(\sv r_1)\bigg({\vec\rho}(\sv r_1|\sv r_2)\cdot{\vec n}(\sv r_2)\bigg) +
\frac{\nu}{2}\bigg({\vec n}(\sv r_1)\cdot{\vec\rho}(\sv r_1|\sv r_2)\bigg){\vec n}(\sv r_2)
$$
$$
- \frac{\nu}{4}\bigg({\vec n}(\sv r_1) + {\vec n}(\sv r_2)\bigg)^2{\vec\rho}(\sv r_1|\sv r_2) + 
\frac{i\nu}{2}\bigg({\vec n}(\sv r_1)\times{\vec n}(\sv r_2)\bigg)\rho_0(\sv r_1|\sv r_2).
$$
Any spatial dependence of $\vec n$ evidently leads to a coupling between $\rho_0$ and $\vec\rho$.
One may proceed in Eq.(\ref{5}) to the Wigner representation. Provided the spatial scale of ${\vec n}(\sv r)$ is greater than the coherence length (the typical width of $\rho_0(\sv r_1|\sv r_2)$ and ${\vec\rho}(\sv r_1|\sv r_2)$ as functions of $|{\sv r_1} - {\sv r_2}|$), the corresponding equations take the following form:
$$
\partial_t\rho_0({\sv r}, {\sv p}) + \bigg(\frac{\sv p}{m}\cdot\nabla\bigg)\rho_0({\sv r}, {\sv p})
$$
$$
= \frac{\nu}{4}\bigg[\bigg(\nabla_i{\vec n}(\sv r)\bigg)\cdot
\bigg(\nabla_j{\vec n}(\sv r)\bigg)\bigg]\partial_{p_i}\partial_{p_j}\rho_0({\sv r}, {\sv p}) +
\frac{\nu}{2}\bigg[\bigg(\nabla_i{\vec n}(\sv r)\bigg)\times
{\vec n}(\sv r)\bigg)\bigg]\cdot\partial_{p_i}{\vec\rho}({\sv r}, {\sv p}),
$$
\begin{equation}
\label{6}
\end{equation}
$$
\partial_t{\vec\rho}({\sv r}, {\sv p}) + \bigg(\frac{\sv p}{m}\cdot\nabla\bigg){\vec\rho}({\sv r}, {\sv p}) = \nu{\vec n}(\sv r)\bigg({\vec n}(\sv r)\cdot{\vec\rho}({\sv r}, {\sv p})\bigg) - \nu{\vec\rho}({\sv r}, {\sv p})
$$
$$
-\frac{\nu}{4}\bigg[\bigg(\nabla_i{\vec n}(\sv r)\bigg)\cdot
\bigg(\nabla_j{\vec n}(\sv r)\bigg)\bigg]\partial_{p_i}\partial_{p_j}{\vec\rho}({\sv r}, {\sv p}) +
\frac{\nu}{4}\bigg(\nabla_i{\vec n}(\sv r)\bigg)\bigg[\partial_{p_i}\partial_{p_j}{\vec\rho}({\sv r}, {\sv p})\cdot\nabla_j{\vec n}(\sv r)\bigg]
$$
$$
- \frac{\nu}{2}\bigg[\bigg(\nabla_i{\vec n}(\sv r)\bigg)\times
{\vec n}(\sv r)\bigg]\partial_{p_i}\rho_0({\sv r}, {\sv p}).
$$
The last term in the RHS of Eq.(\ref{6}) gives an effective force:
\begin{equation}
\frac{d}{dt}\langle p_i\rangle \equiv \frac{d}{dt}\int p_i\rho_0({\sv r}, {\sv p})d^3pd^3r = 
-\frac{\nu}{2}\int\bigg[\bigg(\nabla_i{\vec n}(\sv r)\bigg)\times{\vec n}(\sv r)\bigg]{\vec\rho}({\sv r})d^3r,
\label{7}
\end{equation}
where ${\vec\rho}({\sv r}) \equiv \int {\vec\rho}({\sv r}, {\sv p})d^3p$ is the spatial density of orientation. We see that Eq.(\ref{7}) implies a ``screwing'' of particles into the spin-measuring medium, so that oppositely oriented particles are subjected to opposite forces. Consequently, there will evidently take place a spin separation of particles in an initially prepared unoriented state. The direction of separation in every $\sv r$-point is defined by the double-vector field $\bigg(\nabla_i{\vec n}(\sv r)\bigg)\times{\vec n}(\sv r)$, which is vector both in $\sv r$-space and in the internal spin space. It follows from the last term in the second equation of (\ref{6}) that the same double-vector field is a source for the orientation flux density ${\vec j}_i(\sv r) \equiv \int p_i{\vec\rho}({\sv r}, {\sv p})d^3p$:
\begin{equation}
\partial_t{\vec j}_i(\sv r) = \ldots + \frac{\nu}{2}\bigg[\bigg(\nabla_i{\vec n}(\sv r)\bigg)\times{\vec n}(\sv r)\bigg]\rho_0({\sv r}).
\label{8}
\end{equation}
 	
	Let us consider now an evident but important implication of the pointer basis spatial dependence. Assume that the decoherence rate $\nu$ is extremely large so that the environment instantly suppresses the corresponding off-diagonal elements of the density matrix:
\begin{equation}
\hat{\varrho}({\sv r_1}|{\sv r_2}) \cong |{\vec n}(\sv r_1)\rangle
\varrho_{+}({\sv r_1}|{\sv r_2})\langle{\vec n}(\sv r_2)| + |-{\vec n}(\sv r_1)\rangle
\varrho_{-}({\sv r_1}|{\sv r_2})\langle -{\vec n}(\sv r_2)|.
\label{9}
\end{equation} 
One may say that the environment produces {\it local superselection sectors} ${\cal H}_{+}(\sv r)$ and ${\cal H}_{-}(\sv r)$ in the spin Hilbert space $\cal H$, i.e. 
${\cal H} = {\cal H}_{+}(\sv r)\oplus{\cal H}_{-}(\sv r)$ (${\cal H}_{\pm}(\sv r) = \{|\pm{\vec n}(\sv r)\rangle\}$). Substituting (\ref{9}) into Eq.(\ref{3}), we get
$$
\partial_t\varrho_{s}({\sv r_1}|{\sv r_2}) = \frac{i}{2m}\bigg({\sv \nabla}_{\sv r_1} + 
i{\sv A}_{s}(\sv r_1)\bigg)^2\varrho_{s}({\sv r_1}|{\sv r_2}) - 
\frac{i}{2m}\bigg({\sv \nabla}_{\sv r_2} - 
i{\sv A}_{s}(\sv r_2)\bigg)^2\varrho_{s}({\sv r_1}|{\sv r_2})
$$
\begin{equation}
-\frac{i}{2m}\bigg[\bigg({\sv A}_{s\overline{s}}(\sv r_1)\cdot{\sv A}_{\overline{s}s}(\sv r_1)\bigg) - 
\bigg({\sv A}_{s\overline{s}}(\sv r_2)\cdot{\sv A}_{\overline{s}s}(\sv r_2)\bigg)\bigg]\varrho_{s}({\sv r_1}|{\sv r_2}),
\label{10}
\end{equation}
where $s = +, -$, $\overline{s} \equiv -s$, and
$$
{\sv A}_{\pm}(\sv r) \equiv -i\langle\pm{\vec n}(\sv r)|{\sv \nabla}|\pm{\vec n}(\sv r)\rangle = -{\sv A}_{\mp}(\sv r),
$$
\begin{equation}
\label{11}
\end{equation}
$$
{\sv A}_{+-}(\sv r) \equiv -i\langle{\vec n}(\sv r)|{\sv \nabla}|-{\vec n}(\sv r)\rangle = {\sv A}_{-+}^{*}(\sv r).
$$
This is purely dynamic evolution of ``charged'' particles - the sign of the charge is given by $s$ - in the gauge field of the vector-potential ${\sv A}_{s}(\sv r)$ which includes the charge as a factor. It is remarkable that the locality of superselection rules causes an additional ``gravitational'' potential in Eq.(\ref{10}), $|{\sv A}_{+-}(\sv r)|^2/2m$, which is charge-independent. Note that if the local pointer basis $|\pm{\vec n}(\sv r)\rangle$ depends also on time, extra terms
\begin{equation}
-i\bigg(\varphi_{s}(\sv r_1) - \varphi_{s}(\sv r_2)\bigg)\varrho_{s}({\sv r_1}|{\sv r_2})
\label{12}
\end{equation}   
appear in the RHS of Eq.(\ref{10}), where
\begin{equation}
\varphi_{\pm}(\sv r) \equiv -i\langle\pm{\vec n}(\sv r)|\partial_t|\pm{\vec n}(\sv r)\rangle = - \varphi_{\mp}(\sv r)
\label{13}
\end{equation}
is the analog of the electric potential (the 0-th component of the 4-vector potential). Ordinary $U(1)$-gauge transformations of the vector potential are induced, as follows from Eq.(\ref{11}), by local phase transformations of the pointer basis elements. The ``gravitational'' potential stays unchanged under this transformation.

 %Inserted by TeXtelmExtel

	Resuming, we can say that the spatial dependence of the pointer spin basis does produce effective forces. In the limit of great decoherence rate $\nu$ these forces get purely ``electrodynamic'' form. Hence, the case of finite $\nu$ can be considered as an ``electrodynamics'' with not prohibited interference of positive and negative charges. The dynamic effect of local pointer basis has common neither with potential forces nor with friction-like dissipative forces. The force (\ref{7}) has a peculiar information origin since it is stipulated by measurement-created correlations between the particles' spin state and environment.

 %Inserted by TeXtelmExtel

	A natural question arises as whether there exist real systems demonstrating such forces. In this relation the phenomenon of coherent population trapping [10-13] comes to mind. Given an atom with rapid spontaneous decay into the degenerate ground state and resonant radiation with spatially dependent polarization, one gets the situation which allows reasoning in terms of the space $\cal H$ spanned by the sublevels of the ground state. There are the so-called ``dark'' states in $\cal H$, which do not interact with the radiation. Other states in $\cal H$, orthogonal to the dark states, are called ``bright'' states and are almost empty under the stationary conditions. Dark and bright states form an effective pointer basis. The basis is $\sv r$-dependent due to the mentioned $\sv r$-dependence of light polarization. Hence, the situation is similar to the above considered limit $\nu \longrightarrow \infty$ (Eq.(\ref{10})).

 %Inserted by TeXtelmExtel

	The question is open of how long the similarity between electrodynamic forces and those induced by environment-induced local superselection rules can be spreaded. In particular, it is interesting to consider the counterpart of Aharonov-Bohm effect. Up to my knowledge this issue has not been raised in the mentioned theory of coherent population trapping and subrecoil laser cooling (probably because of complicated geometry of corresponding experiments). The peculiarities of decoherence-induced Aharonov-Bohm effect will be considered elsewhere.\\

 %Inserted by TeXtelmExtel

 %Inserted by TeXtelmExtel

 %Inserted by TeXtelmExtel

 %Inserted by TeXtelmExtel

 %Inserted by TeXtelmExtel

{\bf Acknowledgements}

 %Inserted by TeXtelmExtel

	The author is deeply indebted to Dr. P.Chapovsky, Prof. A.Tumaikin, Drs. A.Taichenachev and V.Yudin for valuable remarks and discussions. Partial support from the Russian Foundation for Basic Research (grant N 98-03-33124a) and Federal Program "Integration" (project N 274) is acknowledged.

 %Inserted by TeXtelmExtel

%\newpage

\end{document}